\documentstyle[12pt,aasms4]{article}
\def\gapprox{{\ _>\atop{^\sim}}}     

\def\Hb{H$\beta$}
\def\Hg{H$\gamma$}
\def\Hd{H$\delta$}
\def\He{H$\epsilon$}
\def\Hx{H$\xi$}

\def\m{$M_\odot$}
\def\l{$\lambda$}

\def\gr{$\gamma$--ray}

\def\Ni{$^{56}$Ni}

\begin{document}

\title{    \makebox[1in]{\ \ }         \\ \makebox[1in]{\ \ }        \\
           \makebox[1in]{\ \ }    \\
         The Luminous Type~Ic SN 1992ar at z=0.145}
\author{ Alejandro  Clocchiatti\footnote{aclocchiatti@astro.puc.cl}\altaffilmark{,17},
Mark M. Phillips\footnote{mphillips@noao.edu}\altaffilmark{,18},
Nicholas B. Suntzeff\footnote{nsuntzeff@noao.edu}\altaffilmark{,18},
Massimo DellaValle\footnote{dellavalle@pd.astro.it}\altaffilmark{,19},
Enrico Cappellaro\footnote{ecappellaro@pd.astro.it}\altaffilmark{,20},
Massimo Turatto\footnote{mturatto@pd.astro.it}\altaffilmark{,20},
Mario Hamuy\footnote{mhamuy@as.arizona.edu}\altaffilmark{,21,28},
Roberto Avil\'es\footnote{raviles@tamarugo.cec.uchile.cl}\altaffilmark{,18,26},
Mauricio Navarrete\footnote{mnavarrete@noao.edu}\altaffilmark{,18},
Chris Smith\footnote{csmith@noao.edu}\altaffilmark{,18},
Eric P. Rubenstein\footnote{ericr@astro.yale.edu}\altaffilmark{,18,22,27},
Ricardo Covarrubias\footnote{riccov@ctiowb.ctio.noao.edu}\altaffilmark{,18},
Peter B. Stetson\footnote{peter.stetson@hia.nrc.ca}\altaffilmark{,23},
Jos\'e Maza\footnote{jmaza@das.uchile.cl}\altaffilmark{,24,27},
Adam G. Riess\footnote{ariess@astron.berkeley.edu}\altaffilmark{,25},
\and
Caterina Zanin\footnote{czanin@pd.astro.it}\altaffilmark{,20}
}
\altaffiltext{17}{Departamento de Astronom\'{\i}a y Astrof\'{\i}sica, P. Universidad Cat\'olica, Casilla 104,
       Santiago 22, Chile}
\altaffiltext{18}{ Cerro Tololo Inter-American Observatory, National Optical
Astronomy Observatories, Casilla 603, La Serena, Chile }
\altaffiltext{19}{Department of Astronomy, University of Padova, Vicolo dell'Osservatorio 5, I-35122 Padova, Italy}
\altaffiltext{20}{Osservatorio Astronomico di Padova, vicolo dell'Osservatorio 5, I-35122 Padova, Italy }
\altaffiltext{21}{University of Arizona, Steward Observatory, Tucson, AZ 85721}
\altaffiltext{22}{Department of Astronomy, Yale University, New Haven, CT 06520}
\altaffiltext{23}{Dominion Astrophysical Obs., 5071 West Saanich Rd., RR 5, Victoria, BC V8X 4M6, Canada} 
\altaffiltext{24}{Departamento de Astronom\'{\i}a, Universidad de Chile, Casilla 36-D, Santiago, Chile}
\altaffiltext{25}{Department of Astronomy, University of California, Berkeley, CA 94720}
\altaffiltext{26}{Departamento de F\'{\i}sica, Universidad de Chile, Casilla 487-3, Santiago}
\altaffiltext{27}{Visiting astronomer, CTIO}
\altaffiltext{28}{Based partly on observations obtained at the
Cerro Tololo Inter-American Observatory, a division of the National
Optical Astronomy Observatories, which is operated by the Association
of Universities for Research in Astronomy, Inc., (AURA), under
cooperative agreement with the National Science Foundation.}

\bigskip
\begin{center}
Submitted to {\em Astrophysical Journal}

\end{center}

\begin{abstract}
We present spectroscopic and photometric observations of SN~1992ar,
the more distant SN in the Cal\'an/Tololo Survey.
We compare its spectrum with those of nearby Type~Ia and Ic SNe and conclude
that the latter type is a better match to SN~1992ar.
Using K--corrections based on the spectra of well observed Type~Ic and
Ia SNe we compute different possible rest frame light curves of SN~1992ar
and compare them with those of representative SNe of each type observed in
the nearby universe.
From the photometry and the spectra, we are able to conclude that SN~1992ar
cannot be matched by any known example of a Type~Ia SN.
Even though the data set collected is fairly complete
(one spectrum and 10 photometric points),
it is not possible to decide whether SN~1992ar was a fast
Type~Ic SN, like SN~1994I, or a slow one, like SN 1983V.
The absolute V magnitudes at maximum implied by each of these possibilities
are $-19.2$ and $-20.2$, respectively.
The latter would make SN~1992ar one of the brightest SNe on record.
SN~1992ar, hence, illustrates the problem of contamination faced by the
high~z Type~Ia SNe samples whose luminosity distances are used
to determine the cosmological parameters of the Universe.
We present observational criteria to distinguish the two SN types
when the SiII$\lambda$6355 line is redshifted out of the sensitivity
range of typical CCD detectors, and
discuss the effect that these luminous Type~Ic SNe
would have on the measured cosmological parameters,
if not removed from the High--z Type~Ia SN samples.
\end{abstract}


\section{Introduction} \label{se:int}

SN 1992ar is one of the supernovae
(SNe) discovered by the Cal\'an/Tololo collaboration (Hamuy et al. 1993).
It was found by R. Antezana, on a plate taken with the CTIO Curtis--Schmidt
telescope on UT 1992 July 27.23  (Hamuy \& Maza 1992),
close to a tight group of three anonymous
galaxies (see Figure~\ref{fi:chart}).
The new object was confirmed as a SN 6 days later, through a spectrum taken
with CTIO 4.0m telescope.
Spectra of the nearby galaxies indicated that the group was at a redshift of
$\sim$0.15, which made this the most distant SN discovered during the
Cal\'an/Tololo survey.
Surprisingly, the spectrum of the SN revealed that the object more closely
resembled a Type~Ic event rather than a Type Ia.
Preliminary calibration of a local sequence of standards and point spread
function fitting photometry of the SN showed that SN~1992ar could
be even brighter than a Type~Ia SN, depending on K corrections which were
unknown at the time.
These brightness estimates were in strong contradiction with the
expectation that Type~Ib and Ic SNe (those formerly known as ``Peculiar''
Type Ia SNe) were subluminous with respect to Type Ia.

The belief that Type~Ib and Ic were subluminous with respect to Type Ia
SNe was rooted in both the observational record and theoretical interpretation.
All SNe of these types, so far, had been 1.0--1.5 magnitudes
dimmer than a typical Type~Ia SN.
The fact that they appear always associated with regions of
very recent star formation suggested that they originated in stars of large
main sequence mass which had lost the outer envelope and, hence,
should explode through core collapse like Type~II SNe.
It is not yet clear what is the ``typical'' amount of \Ni\ that a core collapse
explosion ejects.
Although some events produced as much as $\sim$0.3 \m\ (Schmidt et al. 1994),
comparative analysis of their late time light curves suggest that most of them
eject between 0.04 and 0.10 \m\ (Patat et al. 1994).
This, the successful interpretation and modeling of SN~1987A
in terms of the ejection of 0.07 \m\ of \Ni\ (Woosley 1988; 1990), and the modeling of other
Type~II and stripped--envelope SNe in terms of explosions ejecting similar
amounts of \Ni\ (Wheeler et al. 1993, Shigeyama et al. 1994, Woosley et al. 1994,
Woosley et al. 1995, Clocchiatti et al. 1996a) strengthened the case for low
\Ni\ production in typical core collapse SNe.
The extreme brightness of SN~1992ar raises the question, therefore,
of whether
it might actually be a peculiar Type~Ia event which just happens to bear a
close spectroscopic resemblance with a SN Ic.

CCD observations at only three epochs were obtained from CTIO, the latest
one 17 days after discovery.
In addition, very late deep images of the field, when the SN brightness had
faded beyond detection, were taken with the Blanco 4.0m telescope to allow
subtraction of the galaxy background.
Using this restricted database, Kohnenkamp (1995), did a preliminary study of
this SN.
Additional data, however, had been obtained at ESO, where researchers
working under the Key Program on SNe
followed SN~1992ar for more than 120 days after discovery.
The CTIO and ESO data, when combined, provide a well sampled light
curve with good signal-to-noise ratio, spanning nearly 105 days in the rest frame
of the SN.
These data permit a
detailed comparison of SN 1992ar with typical Type~Ia and Ib/c events.
From this comparison, the nature of SN~1992ar can be finally established.

In Section 2 we present the spectrum of SN~1992ar.
We describe the observations, reduction, and the spectrum itself, and
compare it with spectra of nearby, well observed, Type~Ia and Ic SNe.
In Section 3 we describe the CCD photometric observations, the process
used to eliminate the background light, and the calibration of the
photometry in a standard system.
Section 4 of this paper describes how we obtained a photographic magnitude
from the discovery plate.
Section 5 describes the K corrections we used to translate the observed
photometry into the rest frame of the SN.
In Section 6, finally, we reach the point where a light curve of SN~1992ar
can be presented.
In Section 7 we discuss our results and present our conclusions.

\section{The spectrum}

The confirming spectrum of SN~1992ar was obtained at the CTIO 4.0m Blanco
telescope on UT 1992 August 2.34 by M. Hamuy and R. Williams.
The Cassegrain Spectrograph and a 1240x400 Reticon CCD detector were used.
The wavelength range covered was 3248--7500 \AA, with a
dispersion of 3.6 \AA\ pix$^{-1}$ (approximate resolution $\sim$7\AA).
Two exposures of 10 minutes each were obtained at an airmass of
1.07, approximately, bracketed by exposures of HeAr comparison lamps.
Spectra of the three galaxies around the SN (see Fig.~\ref{fi:chart})
and
two flux standard stars of the list of Hamuy et al. (1994) were also obtained.
All the objects were observed through a 2 arcsec slit,
which was not oriented along the parallactic angle.
The relative flux calibration of the spectrum is, hence,
uncertain due to light loss by atmospheric dispersion.
The absolute flux calibration is uncertain, as well,
since the night was not photometric.

Image reduction and calibration followed the typical procedures.
The spectrum
images were divided by normalized flat--field images built from several
exposures of the dome illuminated by two light sources with different
spectral distribution to provide
good signal-to-noise ratio in all the spectral range.
Spectra of the SN were extracted and sky subtracted.
Special care was taken in fitting the sky underneath the SN profile,
since this SN appeared superimposed on a complicated background (See
Fig.~\ref{fi:chart}).
The spectra were linearized in
wavelength using the HeAr lamp exposures, corrected by atmospheric
extinction using the mean extinction curve for CTIO, and
calibrated in flux using the sensitivity curve for the night, obtained from the
spectrophotometric standards.
The two final SN spectra were normalized to the same mean value
(multiplying the weakest by a constant), and, finally, combined.
Spectra of the three neighboring galaxies were obtained and reduced in the
same way.

The spectrum of the SN
is plotted both in Figures \ref{fi:spec1}, and \ref{fi:spec2},
and the spectra of the neighboring galaxies are given in Figure~\ref{fi:specg}.
The redshifts for each galaxy, obtained from the average of several emission
and/or absorption lines, are, 0.1452, 0.1450, and 0.1462, for galaxies
A, B and C, respectively.
They agree to within $\pm$360 km s$^{-1}$, confirming that the galaxies are
physically associated.
Note that narrow, weak [OII]~\l3727, \Hb, and [OIII]~\l5007 emission
lines at z$=$0.146 are also present in the SN spectrum, pointing at a
close association with the group.

Although it is not completely clear which of the galaxies was the actual host,
circumstantial evidence suggests SN~1992ar was more closely related
with galaxies A and/or B than with galaxy C.
On the one hand, the projected positions of galaxies A and B are closer to
the position of the SN than that of galaxy C.
On the other, galaxies A and B show the strong emission lines characteristic
of galaxies with starburst (Kennicutt, 1992), and Type~Ic SNe are known to
be strongly associated with massive star forming regions in nearby galaxies
(Van Dyk, Hamuy \& Filippenko, 1996).

Using the B \& V transmission curves of Bessel (1990), we found that the color
of the spectrum of SN~1992ar was B$-$V= 0.85.
It is difficult to attach an error bar to this estimate. 
The relative
spectrophotometry from CTIO spectra is typically outstanding (see for
example Figure~2 in Phillips et al. 1992), with RMS uncertainty
on the order of a few
hundredths of a magnitude, similar to that of CCD photometry.
We expect, however, that this would not be the case for SN~1992ar.
One the one hand, this SN was dimmer than the examples available for
comparison, and the background of the parent galaxies was more complicated
than usual.
On the other, our spectrum is not corrected for
atmospheric dispersion.
Although
it is difficult to size the impact of the background on the spectrophotometry,
an estimate of the atmospheric dispersion can be made.
According to Table~1 of Filippenko (1982)
the separation of the intensity
peaks at the effective wavelengths of the B and V filters is smaller than 0.2
arcseconds (i.e., $<$10\% of the slit width)
if the airmass is $\sim$1.07.
Assuming a Gaussian profile for the seeing and integrating the intensity
profiles inside of the slit,
this separation implies that the flux in B could have
been reduced, at most, by 2\% (i.e. our B$-$V color would be at most 0.02 mag
too red).
To be on the safe side, however, we assume an uncertainty of $\pm 0.2$ mag,
about 5 times larger than the uncertainty in the photometry
taken the day before, and 10 times larger than the expected systematic error
due to atmospheric dispersion alone.

Finally, an estimate of the reddening from foreground interstellar matter 
can be made from the absence of NaI~D lines.
According to Burstein \& Heiles (1984) the Galactic extinction in the
direction of SN~1992ar is negligible.
Regarding interstellar matter in the parent galaxy,
the noise peaks in the region of the NaI~D lines at the redshift of
SN~1992ar imply an equivalent width smaller than $\sim$0.5 \AA.
In the absence of a better solution, this upper limit can be used
to estimate an upper limit to the color
excess.
Clocchiatti et al. (1995) find that, in the case of SN~1993J,
$E_{(B-V)}/EW = 0.21$ mag \AA$^{-1}$,
which would imply in the case of SN~1992ar a maximum E(B$-$V) of
$\sim$0.1, or $A_{\rm V,max} \sim 0.3$ mag.
We will neglect in what follows any correction
for reddening.
However, it should be kept in mind that our results could still
be consistent with modest amounts of foreground extinction, up
to a limit of $\sim$0.3 mag in V.

The spectrum of SN~1992ar is compared in Fig.~\ref{fi:spec1} with spectra of
the ``normal'' Type~Ia SN~1989B and the peculiar Type~Ia SN 1991T.
The phases of the 1989B and 1991T spectra were chosen to be consistent with the
phase implied for our spectrum of 1992ar (were it a SN Ia) by a spectroscopic
matching technique (see \S 6.1).
We can see that the similarities of the SN~1992ar spectrum with those of Type~Ia
SNe are, mainly, the broad features to the blue of $\sim$5300\AA.
These absorption features and P~Cygni profiles are, from red to blue, due to
FeII \l5169,
FeII \l5018,
FeII ``\l4555'',
FeII ``\l4274'',
and
CaII H \& K.
Since they are present in the spectra of SNe of all types, including Type~II, 
their potential to differentiate a SN from another is very limited.
Features of CaII, FeII, and NaI, which are common to Type~Ia and Ic SNe, and
the SiII \l6355 line, which, although weaker than in Type~Ia,  also appears in
Type~Ic SNe, are labeled with lower case letters.

Even though the similarities in this spectral region are clear, there
are a few differences worth noting.
The features of SN~1992ar are shallower and broader, and the spectra of Type~Ia
SNe show narrow cores in the absorption lines.
These differences are particularly noticeable between 3900 and 4200~\AA, and
between 4700 and 5100~\AA.
In the former region, the main difference is made by the feature labeled with
letter A, which corresponds to SiII \l4130.
In the latter region, there is an additional feature, labeled with letter B,
which corresponds to SiII \l5051.

It is, however, in the region to the red of $\sim$5200~\AA\  that the distinct
character of the spectrum of SN 1992ar becomes clear.
First, the strong SiII \l6355 line, centered between 6100 and 6200~\AA\ 
in SN~1989B and 1991T, is {absent} in SN~1992ar.
Second, the much weaker but also distinctly present absorption centered near
5500\AA\  in the Type~Ia SNe, typically attributed to SII~\l5612 and
SII~\l5654 (labeled with letter D in the figure), is also absent.
There is another, much weaker, feature about 200 \AA\  to the blue, labeled
with letter C in the figure, which corresponds to SII \l5468.
The signal-to-noise ratio of the spectrum of SN~1992ar is not good enough to decide
whether this small feature is present or not.

Strong SiII lines and clearly detectable SII lines are characteristic
features of Type~Ia SNe.
The fact that SN~1992ar does not show them is a clear indication that, if it
were a Type~Ia SN, it would be even more spectroscopically peculiar than
SN~1991T.

Consistent with this, Fig.~\ref{fi:spec2} shows that there are no major
differences between the spectra of typical
Type~Ic SNe 1983V or 1987M and that of SN~1992ar.
With the exception of the CaII H \& K lines,
the spectrum of SN~1983V is remarkably similar to that of SN~1992ar.
However, since Type~Ic SNe do show differences in the strength of the CaII
lines, the weaker spectral features are not a major concern.
Paradoxically, one of the notable differences in this figure is, again, the
SiII \l6355 line.
Although it is clear that the spectra shown Fig.~\ref{fi:spec2} are type Ic,
one sees that this line was stronger in SN~1987M than in the other two.

\section{CCD Photometry}

We have collected 10 CCD images of SN~1992ar taken from telescopes at CTIO and
ESO.
A description of the dates, telescopes used, observers, and other
relevant information is given in Table~\ref{ta:CCDobs}.
The images, taken through standard Johnson B, V (Johnson 1955),
and Cousins R (Cousins 1976a, 1976b, 1978)
filters were reduced following standard procedures.
The CCD pedestal was corrected using its overscan region,
any remainder 2-dimensional pattern was removed using bias images when
necessary.
No correction for dark current was required.
Finally, pixel to pixel sensitivity variations were removed using median
dome or sky flat field images.

SN~1992ar appeared on a complicated background (see Fig.~\ref{fi:chart}).
The nearby galaxies 1 and 2 mean that 
the sky near the SN is bright and has a large gradient.
In addition, the SN is centered
on a valley of the brightness distribution.
This configuration makes both aperture and PSF fitting
photometry bad prospects.

In order to do unbiased photometry, we resorted to eliminating the light of
the nearby galaxies by taking
very late-epoch images of the region, with good signal-to-noise ratio.
Information on these images,
which we refer to in what follows as the final epoch images, is
given in Table~\ref{ta:FEImages}.
We used the image matching algorithms described by Phillips \& Davis (1995),
proceeding in detail as follows:
(a)
We shifted and rotated the final epoch image so that the centers of the stars
were in coincidence with those of the image with the SN;
(b) we found the difference kernel that, when convolved with the image of the
narrowest PSF (usually this is the final epoch image),
would degrade it to match the PSF of the other image, and
convolved the image with better seeing with that kernel;
(c) we found the linear transformation in intensities to make the final epoch
image match the intensities of the image with the SN, and applied that
transformation to the final epoch image;
(e) we subtracted the region around the SN of the transformed final epoch image
from the image with the SN.
After this procedure we found that the background of the galaxies near the
SN had been removed, and that the sky underneath the SN was flat.

We did relative photometry of the SN on
the images with the background subtracted.
Using the tasks of the package DAOPHOT in IRAF\footnote{ The Image Reduction
and Analysis Facility (IRAF) is maintained and distributed by the Association
of Universities for Research in Astronomy Inc., under a cooperative agreement
with the National Science Foundation.}
we built a mean PSF for each frame from the profiles of the stars
in the field
and found the instrumental magnitudes of the local sequence of
standards and the SN by fitting this PSF function.

The sequence of local standard stars indicated in Fig.~\ref{fi:chart}
was calibrated in Johnson B and V colors 
using observations taken with CTIO 0.9m telescope on August 22, 1993.
On that night, images of the region of SN 1992ar were taken together with
standard stars of the lists of Graham (1982), Menzies et al. (1989),
and Landolt (1992).
The zero points, extinction coefficients, and color terms were fitted
to find the photometric solution for the night, and
the B and V magnitudes given in Table~\ref{ta:STD}
were computed from this solution.

We did not have a photometric night with
enough standard star observations to compute all the coefficients in the
photometric solution for the Cousins R filter.
The final epoch image, however, was taken on a photometric night and our set
of images included three standard stars of the list of Landolt (1992).
We used these stars, a mean extinction coefficient for La Silla
from the database of the Photometry Group of the Geneva Observatory 
(Burki et al. 1995), and a typical color term for the instrument used
(Turatto, 1996), to find the zero point of the calibration in R.
Due to the characteristics of this solution, there could be a systematic error
in the calibrated R magnitude for the local sequence of standards.
It must be small, however, of the order of the uncertainties given in
Table~\ref{ta:CCDobs}, since the two color diagram of the local standard
sequence looks reasonable.
We decided to give the R values out of completeness.
In the event that these particular observations of SN~1992ar become important
(to estimate the foreground extinction by comparing the late time V-R
color with those of future Type~Ic SNe, for example) the
sequence can be recalibrated and a correction to the values given
in Table~\ref{ta:CCDobs} could be computed.

Using the instrumental magnitudes of the stars in the frames with the SN
and their B, V, and R magnitudes, we computed  a zero point for each
frame and applied it to the instrumental magnitude of the SN.
The color terms for all the combinations of CCD and instruments used is
small, so color term corrections to the magnitude of the SN relative
to the local standard stars are expected to be smaller than
the photometric uncertainty.
Consistent with this, we found the zero points to be independent of the
color of the local sequence stars.
The magnitudes of the SN, calibrated in this way,
are given in Table~\ref{ta:CCDobs}.

\section{Photographic Photometry}

As it will become clear in the next section,
to find the nature of SN 1992ar requires us to use all possible
data at hand, including its discovery plate.
A complication in this regard is that the search plates of the Cal\'an/Tololo
survey were taken without a filter.
However, the standard Kodak IIa0 emulsion, with a red cut--off very similar to
that of the Johnson B band, was employed.
This, together with the fairly red color of SN~1992ar, suggest that it should be
possible to get a good estimate of its B magnitude from the photographic
magnitude.

There were three Schmidt plates in the database of Cerro
Cal\'an Observatory with
images of the region of SN~1992ar (Wischnjewsky 1997).
One of them is the first epoch image, which was taken on June 26.5 1992,
the second is the discovery image, taken on July 27.3 1992, the third is
an additional plate taken on August 2.5 1992.
Ideally, these plates could have been used
to digitize the region of the SN and sequence of local
photometric standards in all the plates, transform the photographic
density to an approximate intensity scale, use the plate of June 26.5 to
estimate the background underneath the SN and subtract it from the other two
plates,
compute magnitudes for the local standards and the SN, analyze
the correlation between the instrumental photographic magnitudes and the
calibrated magnitudes for the standard stars in all plates,
and use this correlation to
estimate the photographic magnitude of the SN in a standard passband.
Since by August 2.5 the light curve is known from CCD observations, the
calibration of the August plate would be useful to check the accuracy of the
photographic photometry.

We were able to accomplish this project only partially.
On the one hand, we found that
the photographic density of the sky in the region of the SN was only
marginally detectable 
in the June image (negative observation) and, hence, dominated by noise.
This made it useless to model the background of the SN.
In addition, the seeing of the plate of August 2.5 was
worse than that of the discovery plate.
Since the SN was fainter on August 2.5 than on July 27.3,
the plate of August provides only a marginal
detection of the SN, not good enough to set a meaningful
constraint on the precision of the photographic photometry.

We decided, then, to use our final epoch CCD images to
model the background of the SN in the digitized discovery plate, and
desist in our desire to check the accuracy of the
photographic photometry.
We proceeded as follows.
Building upon the suggestions of Stetson (1979), especially his equations
9--11 we took the photographic density to be proportional to the
intensity, an approximation which holds true in the limit
of stellar images of low intensity above the background.
We transformed our final epoch CCD images in the B filter to the pixel scale
of the plate, studied the correlation (pixel by pixel) of the intensity
recorded by the CCD and the photographic density in regions of low density,
and found that they were actually consistent with a linear transformation.
We found the linear transformation that best fit the correlation, and
applied it to the CCD image.
We then took a small region of the transformed CCD image, centered on the SN
position, and subtracted the transformed CCD intensity distribution of this
region from the photographic density.
We found that the sky around the SN was acceptably
well subtracted, leaving residuals only near the cores of the nearby
galaxies (where the assumption of CCD intensity proportional to photographic
density was expected to break down). 

We used this final image to compute the
photographic index of the local standard sequence and SN.
We did not have enough stars to model the photographic PSF shape and correct
for saturation at high photographic density (as in Stetson, 1979).
We resorted, therefore, to computing our photographic index using aperture photometry.
Using 7 stars imaged both in the CCD and the plate,
we found a tight linear correlation between our photographic index and
the B magnitude of the stars in the local sequence.
The scatter of the correlation,
which spans 2.6 magnitudes in the photographic index, is 0.05 mag.
The residuals of the fit, in addition, display a small but statistically
significant color term.
From the correlation and color term obtained, the 
B magnitude of SN 1992ar on the discovery plate is given by,
\begin{equation} \label{eq:Bphot}
{\rm B} = 20.70 \pm 0.17 - (0.183\pm 0.032) \,{\rm (B - V) }
\end{equation}
The color term in this equation is the same, within the uncertainty, as
those found by Pierce \& Jacoby  (1995) and Arp (1961), in transforming
magnitudes from IIaO plates to B.

The uncertainty given in equation ~\ref{eq:Bphot} results from combining
in quadrature the uncertainty in
the instrumental magnitude of the SN and the RMS of the
transformation of the stars in the local sequence.
There is an additional uncertainty, that we cannot evaluate,
because the SN is fainter than the faintest star in
the local sequence of the plate. The linear correlation was
{\em extrapolated} $\sim 0.8$ magnitudes to reach
the photographic index of the SN.
It is clear, in addition,
that in order to derive a B magnitude we need to estimate the (B$-$V) color
of the SN.
This estimate will introduce another source of uncertainty.

\section{K Corrections}

At a redshift of z$=$0.145, it is important to consider the modification of the flux
distribution due to the Doppler effect, and apply the corresponding K corrections
to the observed magnitudes, if comparison with SNe observed in the nearby
Universe is to be done.
We have computed K corrections for the B and V magnitudes of
SN~1992ar following the usual
prescriptions (see Hamuy et al. 1993, Kim et al. 1996, or Schmidt 1998,
for modern discussions), and using spectra
of SN~1983V (Clocchiatti et al. 1997), SN~1987M (Filippenko, Porter \&
Sargent 1990), and SN~1994I (Filippenko et al. 1995,
Clocchiatti et al. 1996b, Jeffery et al. 1994).
A deeper consideration of K corrections for Type~Ic SNe will be given elsewhere
(Clocchiatti 1999).
It is necessary, however, to give a short discussion here.

Type~Ic SNe present a more complex problem than Type~Ia SNe, when it comes
to applying K corrections.
Consider first that, as emphasized by Hamuy et al. (1993), the K-corrections
for a SN qualitatively behave as a color.
Now,
even when Type~Ia SNe are not as homogeneous as it was thought just a few
years ago, their differences in color evolution do not translate into { major}
differences in the K corrections computed from spectra of different SNe.
Even when intrinsic differences in the SNe increase the scatter in the
K corrections the results are still reasonably
consistent with a unique K correction function
for all SNe, at least at redshifts smaller than z$\sim$0.5 (Hamuy et al. 1993).
Type~Ic SNe, on the other hand, present a much wider range in the speed
of the light curve and color evolution.
Some of them are extremely fast, like SN~1994I (Richmond et al. 1996),
while some others are slow, like SN~1983V (Clocchiatti et al. 1997).
These two in particular
were considerably faster and slower, respectively, than a typical Type~Ia SN.
Accordingly, the K corrections for a Type~Ic SN
are not a unique function
of time but critically depend on the speed of the light curve.
In Figure~\ref{fi:Kcorr} we give the K correction in V for a fast and a slow
Type~Ic SN at
z$=$0.145 as a function of rest frame time since maximum in V, together with
the adopted fitting functions.
The RMS of the interpolation is 0.06 mag,
for fast Ic SNe, and 0.05 for slow Ic SNe.
The RMS around the fit for the K correction in B (not given in
Fig.~\ref{fi:Kcorr}) is 0.06 mag.
We will take these values as a measure of the uncertainty introduced
by the K corrections.

In addition, it was necessary to test the possibility that
SN~1992ar was a Type~Ia SN with a peculiar spectrum.
Lacking anything better,
K corrections based on typical Type~Ia SN spectra were used for this.
We took those given by Hamuy et al. (1993), interpolating their values using
piece$-$wise low degree polynomial functions.
The adopted corrections are also given in Fig.~\ref{fi:Kcorr}.
The RMS around the interpolating function is 0.05 mag in this case.

A different K correction is needed for the B magnitude obtained from the
discovery plate, since the sensitivity function of the telescope plus plate
system is not matched by a standard filter.
We used the sensitivity function computed by Pierce \& Jacoby (1995) for the
Kitt Peak Schmidt telescope plus a IIa0 plate, under the assumption that it
must be similar to that of the Tololo Curtis-Schmidt telescope with the same
kind of plate.
An additional problem in this case is that this sensitivity function is very
blue, and it is difficult to find SNe spectra blue enough to cover
this passband.
According with the different SN type and phase we were trying to replicate,
we used literature spectra of the Type~Ic SN~1994I and the Type~Ia SN~1992A,
together with their HST spectra (obtained by the SINS collaboration,
Jeffery et al. 1994), to compute these K--corrections.

The K correction functions displayed in Fig.~\ref{fi:Kcorr} illustrate the
problem of using the same standard filter in the rest and redshifted frames.
Not only are the resulting K corrections function of time, but the time scales
of the variations {\em and their amplitudes},
are similar to those which differentiate one type of SNe light curve
from the others, even at the comparatively small redshift of SN~1992ar.
This effect emphasizes the importance of observing high redshift SNe in
filters that provide a better match to the wavelength range where the light
curves used for comparison were observed.
These matches can be achieved via the coincidence of effective
wavelengths of standard filters for a given redshift (Kim, Goobar \& Perlmuter
1996), or the construction of new photometric systems which are the
redshifted counterparts of the standard one (Schmidt et al. 1998).

\section{The Light Curve}

Undilating the time scale using the redshift measured from the emission lines
in the nearby galaxies, it is clear that the observed photometry is consistent
with that of a SN (see Figure~\ref{fi:lcexample}).
It is clear, also, that the light curve does not correspond to that of a
typical Type~II plateau event.
Finally, it would seem that the confirming spectrum was
taken close to maximum light in V.

The strong time variation of the K corrections, however, as illustrated in
Fig.~\ref{fi:lcexample}, implies that obtaining the correct final light
curve requires a decision as to which of the functions plotted in
Fig.~\ref{fi:Kcorr} is the correct K correction to apply.
The analysis presented in \S~2 supports the original identification of
SN~1992ar as a Type~Ic SN (Hamuy \& Phillips 1992).
The SN is, however, surprisingly bright for being a Ic.
If H$_0 = 65$ km s$^{-1}$ and q$_0 = 0$, then the distance modulus is
39.28 magnitudes, and the first CCD point implies
M$_{\rm V} = -19.2$ mag, without applying any K correction (which could make
the SN brighter), or residual foreground extinction (which {\em will} make the
SN brighter)
\footnote{We take H$_0 = 65$ km s$^{-1}$ here and throughout the paper in
order to compare the absolute brightness of SN 1992ar with those of the other
Cal\'an/Tololo SNe given by Hamuy et al. (1996).}.

According to Hamuy et al. (1996),
the absolute V magnitudes of Type Ia SNe range between $-$19.5 and $-$18.5,
approximately, for the same value of H$_0$.
SN~1992ar therefore compares in brightness with the
brightest Type~Ia SNe in the Cal\'an/Tololo survey.
Other Type~Ic SNe observed near maximum are clearly subluminous
with respect to typical SNe Ia.
SN~1983V (Clocchiatti et al. 1997), a slow Type~Ic SN,
reached M$_{\rm V} =$ -18.1 $\pm$0.2, with the uncertainty dominated by
the reddening estimate.
SN 1994I (Richmond et al. 1996), reached M$_{\rm V} =$ -18.1 $\pm0.6$, with
the big uncertainty due to a large foreground extinction.
The fact that SN~1992ar was as much as a magnitude
brighter than these two well observed nearby Type~Ic SNe
raises the question of whether it
could have been a Type~Ia SN with a peculiar spectrum.
We analyze this possibility in the following subsection.

\subsection{The case for a Type~Ia SN}

The spectrum of SN~1992ar, as we discussed before, is more consistent with
the spectrum of a Type~Ic SNe than with that of a Type~Ia.
This is so, however, mainly because of the missing Si~II \l 6355 line.
There is at least one example of a Type~Ia SN which
showed a very weak Si~II \l 6355 line: the luminous, slow--declining, and
spectroscopically peculiar SN~1991T (Filippenko et al. 1992,
Phillips et al. 1992).
It is important to note that 
the peculiar character of the spectrum of SN~1991T,
and the weakness of the Si~II \l 6355
line, were more evident in the pre--maximum stage.
Shortly after maximum, the spectrum of SN~1991T became more similar to the
typical spectrum of a SN Ia (note that the spectrum of SN~1991T plotted in
Fig~\ref{fi:spec1} already shows the Si~II \l 6355 line).
Hence, if SN~1992ar were a 1991T--like event, the spectrum and the photometry
points taken shortly before and after it, should most likely
correspond to a maximum or premaximum stage.

To test whether SN~1992ar could have been a spectroscopically peculiar
Type~Ia SN we compared the photometry of Table~\ref{ta:CCDobs} with typical
light curves of Type Ia SN of different speed classes (Hamuy et al. 1996a).
We proceeded as follows.
We assumed a reasonable phase for the first point (between $-$6 and 12 days
after maximum in V), computed the K correction for each observed phase,
corrected the CCD photometry using these K--corrections and
fit the resulting data to one of the template SN~Ia curves given by Hamuy et
al. (1996b).
We then computed the RMS of the fit,
changed the phase by one day, and iterated the process until finding the phase
that provided the minimal RMS.
The results of this exercise for the light curves of SN~1992bc, a bright
Type~Ia SN with a slow decaying light curve similar to that of SN~1991T,
and SN~1992A, a fairly typical Type~Ia SN, are displayed in
Figure~\ref{fi:LCIa1}.
Usage of a fast decaying Type~Ia light curve, typical of underluminous
SNe like SN~1991bg, did not provide a better match to the exponential tail;
since it is clear that SN~1992ar was not a low--luminosity event, we do not
show those fits.

The fits in Figure~\ref{fi:LCIa1} show that, for reasonable values of the
phase, the light curves of Type~Ia SNe provide a poor match to the observed
points.
In particular,
it is not possible to obtain a reasonable fit of either the decay
after maximum or the exponential tail 
with the light curve of a slow decaying Type~Ia SN 
(top panel in Figure~\ref{fi:LCIa1}).
The fit is particularly bad if a
a negative phase for the spectrum,
taken about one day after the first CCD point, is assumed.
The absolute magnitude at maximum with this fit would be M$_{\rm V} =$ -19.08,
(H$_0 = 65$ km s$^{-1}$, q$_0 = 0$), on the dim side for a slow--declining
Type~Ia SN.
Using the light curve of a faster--declining Type~Ia SN, like SN~1992A (bottom
panel in Figure~\ref{fi:LCIa1}), the decay after maximum is reasonably well
matched and the absolute magnitude at maximum results M$_{\rm V} =$ -19.2,
similar to those of the comparable Type~Ia SNe in the Cal\'an/Tololo survey.
However, the well observed exponential tail is overluminous by
$\sim$ 0.6 mag, and allows us to reject this fit.

The only way to obtain a decent fit using the light curve
and K corrections of a Type~Ia SN is by releasing the requirement that the
phase of the first CCD point is ``reasonable,'' and by
assuming that it corresponds
to $\sim$17 days after V maximum (see Figure~\ref{fi:LCIa2}).
It is at this point that an estimate of the brightness at the time of
discovery, based on the discovery plate, could be useful.
Using equation~\ref{eq:Bphot} plus an estimate of the color of the SN
at the time of discovery or the color evolution from the time of the
spectrum and the color measured from the spectrum,
we obtained the V magnitudes given in Table~\ref{ta:mphot}.
The color of the SN obtained in this second case is too blue for a SN
Ia 12 days after maximum in V.
As we see in Figure~\ref{fi:LCIa2}), however,
both of the V estimates are consistent
with the light curve, given the large uncertainties involved.

Even when the interpretation of SN~1992ar as a typical Type~Ia SN caught at
late times cannot be ruled out with the sole argument of its light curve,
it is questionable on two grounds.
First, the maximum of SN~1992ar would have occurred around
July 12, 1992, at V$\sim 19.1$.
This implies M$_{\rm V} =$ $-$20.2 (H$_0 = 65$ km s$^{-1}$ Mpc$^{-1}$, q$_0=0$),
more than 0.6 mag brighter than the brightest Type~Ia SN in the
Cal\'an/Tololo survey.
Second, the phase of the spectrum would be $\sim$18 days after V maximum,
which is inconsistent with the spectral features seen in Fig.~\ref{fi:spec1}.
Using a more quantitative approach, spectroscopic dating of the spectrum
as described by Riess et al. (1997), indicates that the phase of this spectrum
(if it were the spectrum a Type~Ia SN) was $6 \pm 2$ days after V maximum, in
gross contradiction with the phase implied by this particular
fit of the light curve.
These considerations are a strong indication that the photometric evolution
and spectral age of SN~1992ar cannot be reasonably matched
by any known examples of Type~Ia SN light curves  and spectra.

\subsection{The case for a Type~Ic SN}
\label{subse:IcSN}

Using the light curves and
K--corrections of { slow} and { fast} Type Ic SNe,
we repeated the procedure of assuming a phase for the first photometric
point,
computing K--corrections, fitting the light curve for a template,
computing of the RMS, and iterating on the
phase until the RMS was minimal.
As a model of the light curve of a { fast} Type~Ic SN we took
SN~1994I (Richmond et al. 1996) and as an example of a
{ slow} Type~Ic SN we chose the light curve of SN~1993J (Richmond et al.
1994). The latter SN was a transitional Type~II, which showed strong He~I lines
starting $\sim$20 days after explosion.
The second maximum of its light curve and early exponential decay are
a good match to the light curves of the slow Type~Ic SNe~1983V (Clocchiatti
et al. 1997) and 1990B (Clocchiatti et al. 1999).
These kind of light curves are probably representative of a photometric group
that includes SNe of spectral types Ib, Ic and II transition to Ib.
The best fitting light curves and phases are displayed in Figure~\ref{fi:LCIc}.

It is clear from the figure, that both of the template light curves
provide a reasonable match to the CCD photometry
within the observational uncertainties
and expected intrinsic scattering of each SN subclass,
to the point that it is not possible to discard either.
The fitting of a fast Ic light curve implies, like in the case of the Ia
templates, that maximum occurred several days before the first photometric
data were obtained (c.f. Fig.~\ref{fi:LCIa2}).
The absolute magnitude implied for maximum would be M$_{\rm V} \sim -20.2$ mag,
again, more than 0.6 mag brighter than the brightest Type Ia SNe
in the Cal\'an/Tololo survey.
In this case, however, the spectrum 
is reasonably similar to that of other Type~Ic SNe
$\sim$8 days after V maximum (the phase implied by the fit) and
thus does not give an argument to discard it.

The fitting of a slow Type~Ic SN is also good (see bottom panel in
Fig.~\ref{fi:LCIc}) , and
implies an absolute magnitude at maximum of M$_{\rm V} =$ -19.26 mag,
similar to the brightest SNe in the Cal\'an/Tololo survey.
It is clear from the figure that 
the goodness of the fit decreases as the phase increases.
This is, however, the expected behavior for stripped--envelope SNe
and is consistent with the usual picture of them originating
in massive stars which have lost their outer layers.
A light curve faster on the tail implies a more rapid decay of the
total \gr\  optical depth and is related with a smaller mass--to--energy
ratio in the ejecta (see Clocchiatti \& Wheeler, 1997, for a simple model).

Again, we can try to decide between this two different interpretation by
resorting to brightness estimates based on the discovery plate.
From the phases provided by the fits, the B$-$V color measured from the
spectrum, and the assumption of a B$-$V color evolution similar to those of
SN~1994I (Richmond et al. 1996), or SN~1983V (Clocchiatti et al. 1997),
we can estimate the color of the SN at the time of the discovery plate
for the cases of a fast and slow Type~Ic SN, respectively.
Another color estimate can be obtained under the assumption that it was
similar to that of either SNe {\em at} the given phase.
With these color estimates and equation~\ref{eq:Bphot},
we can derive its B and hence V, magnitudes.
These estimates are given in Table~\ref{ta:mphot}.
We have these V estimates for each SN type in Fig.~\ref{fi:LCIc}.
As one can see, they are
consistent with both the light curve of a slow Type~Ic SN and
that of a fast one, perhaps slightly better with the latter.

Hence,
surprisingly, the photographic plate with the discovery observation of
SN~1992ar
does not allow us to discriminate between the fast and slow SN Ic templates.
Even when the difference between the two interpretations at the time of
discovery is about 0.8 mag,
the different speed of color and K--correction evolution combine in such a
manner so as to make the plate consistent (within the large uncertainties),
with either hypothesis.

To summarize, therefore, we can only conclude that, if a phase consistent
with that of the spectrum is assumed, the light
curve of SN~1992ar is best fit by those of Type~Ic SNe.
We cannot discern, however, whether it was a fast or a slow one.
The rest frame phases, magnitudes, and K corrections implied by the Ic
SN fits are given in Table~\ref{ta:Icfits}.
The rest frame B$-$V color on the date of the spectrum
($\sim 4$ rest frame days after V maximum for a slow Type~Ic SN, and
$\sim 8$ rest frame days after V maximum for a fast Type~Ic SN)
would have been $\sim$0.62 mag, in both cases, in reasonable
agreement with those
of other Type~Ic SNe at these phases (Clocchiatti et al. 1997, Richmond et al.
1996) when corrections by foreground
extinction (to the other SNe) are considered.

\section{Discussion and Conclusions}

SNe of the Types Ib, Ic, and II-transition to Ib (hereafter referred to as
{\em Transition Type SNe})
display a great degree of heterogeneity both in the appearance of the
spectra and in the photometric behavior.
Typically understood as stars originally massive which have lost their light
element envelopes through episodes of stellar wind and/or mass transfer to
interacting binary companions, this diversity has been taken as indicative of
a large variety of possible configurations of the progenitor star.
Even the three parameter space of two initial stellar masses and their
separation, without consideration of the all the possibilities of the
complex physics involved,
provides many different branches of interacting stellar evolution, and has
the potential of providing a variety of combinations of core and envelope
at the time of core collapse.
Different subsets of these combinations have been proposed to explain SNe
of the spectroscopic types Ib, Ic, II$-$L, and II$-$n.
A handful of these have been studied, with their theoretical
light curves and spectra computed, and shown to match the observed
light curves and spectra of nearby bright SNe like 1983N, 1993J and 1994I
(Woosley et al. 1994); Woosley, Langer \& Weaver 1995; Nomoto et al. 1997).
All of these recent studies assume the
same basic explosion mechanism: The gravitational collapse of an iron core
that creates a shock, produces a last episode of stellar nucleosynthesis and
disrupts the star, ejecting a relatively small amount of \Ni\ ($\sim$0.10 \m).

The extreme luminosity of SN~1992ar introduces yet another dimension of
heterogeneity in this type of explosion.
We have collected in Table~\ref{ta:absmag}, the absolute magnitudes of some
of the best studied Transition Type SNe.
Although all of them were affected by foreground extinction in various
amounts, and this implies a large uncertainty in some cases, it is clear that
SN~1992ar's luminosity is unprecedented.

From the spectrum displayed in Fig.~\ref{fi:spec1}, the mass to energy ratio
of SN~1992ar, as indicated by the square of the expansion velocity,
does not appear to be terribly different from that of SN~1987M or 1983V.
It is not fair to attach a quantitative estimate to the differences
that may exist, however, since the spectra compared
correspond to an early epoch.
Mass to energy ratios are better represented by the expansion
velocities when they reach the asymptotic behavior,
typically at the beginning of the radioactive tail.
The slope of the light curve, on the other hand, under traditional
interpretation of the \gr\  deposition in expanding spherical shells is
also related to the mass to energy ratio.
According to the simple model of Clocchiatti \& Wheeler (1997),
for the same mass--density structure and
effective \gr\  opacity per unit mass,
the difference in slopes some 100 days after maximum indicates that the
mass to energy ratio of SN~1992ar was approximately 1.2 times smaller than
that of SN~1993J.

We can also estimate the amount of \Ni\  powering the light curve of SN~1992ar.
Modeling of the light curve of SN~1993J by different groups, at the distance
given by recent calibration of Cepheid variables in the parent galaxy (M81),
suggests an ejected mass of radioactive \Ni\  from 0.073 \m\ 
(Woosley et al. 1994) up to 0.1 \m\  (Nomoto et al. 1997).
Assuming that both the percentage of energy from radioactive decays
and the bolometric correction at maximum were similar in 
SN~1992ar and SN~1993J,
the mass of \Ni\  ejected in the latter can be obtained using the
absolute magnitudes in Table~\ref{ta:absmag}.
The flux in V of SN~1992ar was about 4 times larger than that of SN~1993J,
and thus the \Ni\  implied for SN~1992ar is between 0.3 \m\ and 0.4 \m.
In the case of a fast Type~Ic SN light curve fit, we can compare again with
SN 1994I.
Iwamoto et al. (1994) estimated an ejected \Ni\  mass of
0.07$^{+0.035}_{-0.025}$ \m.
According with Table~\ref{ta:absmag}, the flux in V of SN~1992ar was 
approximately 7 times larger than that of SN~1994I so a scaling of Iwamoto's
estimate will put the \Ni\  mass of SN~1992ar at $\sim$0.5 $^{+0.25}_{-0.18}$.
Both of
these estimates indicate that the \Ni\  mass ejected by SN~1992ar
should have been comparable to that produced by a fairly
bright Type~Ia SN (Arnett 1996).

A subset of Transition Type SNe, including the slow Type~Ic events,
follows a similar photometric evolution
even though they display a variety of spectra at maximum.
SN~1993J, which is
a good example of the objects in this photometric group,
is well understood in terms of a ``traditional'' core collapse
explosion.
Could one of these explosions produce as much \Ni\  as required by 
the slow Type~Ic SN fit to SN~1992ar?
Or, alternatively, could a core collapse explosion in a $\sim$2\m\  C$+$O star
like the one used by Iwamoto et al. (1994) to match the light curve of SN~1994I,
produce half a solar mass of \Ni ?
The other option is to conclude that there are some stripped envelope SNe which,
contrary to conventional wisdom, do explode via a thermonuclear explosions
(Branch, Nomoto \& Filippenko 1991).
Whether a thermonuclear explosion could give a Ic spectrum
together with the fast decline after maximum
and slow decline on the tail of a slow Type~Ic SN,
{\em or} the fast light curve of a fast Type~Ic SN,
is a question that merits further study.

The existence of luminous Type~Ic SNe such as SN 1992ar has important
implications for the the high--redshift
SNe searches currently under way with the aim of measuring the
cosmological parameters of the universe
(Perlmutter et al. 1998, Schmidt et al. 1998)
because it exemplifies the problem of sample contamination.
SN~1992ar shows that there are SNe which are as bright as a Type~Ia SN and have
a similar light curve near maximum light,
but have different intrinsic
colors, different color evolution, and do not follow the
Light Curve Shape--Luminosity relation used to calibrate the intrinsic
luminosity of a Type~Ia SN.

One of the earliest and clearest indications that SN~1992ar was a peculiar
event was given by its spectrum, which did not display the SiII \l6355 line.
If SN 1992ar had been at a redshift larger than $\sim$0.50, however,
this line would have been Doppler shifted to $\sim$9200~\AA, and the
brightness at maximum would have been around R$\sim$23 mag.
This, together with the poorer sensitivity and typical fringing of most
spectrographs in the far red, would have made it very difficult for
the observers to distinguish on spectroscopic grounds such a SN from a
typical SN~Ia (See Riess et al. 1998, for an example of this problem).
The light curve (which almost surely would have had worse signal-to-noise ratio
and shorter time baseline than those of SN~1992ar)
would have probably been acceptably
fitted with a typical Type~Ia like curve, using
parameters similar to those of the bottom panel in Fig.~\ref{fi:LCIa1}.
Note that this fit implies a phase of $\sim$ 7 days after V maximum for the
spectrum, in good agreement with the spectroscopic age derived by the
method of Riess et al. (1997) for the spectrum of SN~1992ar.
Using the K corrections for Type~Ia SN,
the rest frame color at the time of the spectrum implied by this fit 
($\sim$7 days after V maximum) is B$-$V= 0.6 mag,
somewhat red in comparison with a typical Ia SN $\sim 10$ days after
B maximum.
This red color, however,
would have been interpreted as indicative of foreground extinction,
and the intrinsic brightness of SN~1992ar would have been increased accordingly.
In the end, the luminosity distance derived from this object (if it were
assumed to be a Type~Ia SN) would have been too short.
The possibility exists, therefore,
depending on the signal-to-noise ratio of the observational sets
obtained for a particular object, that the samples of SNe with z$\gapprox$0.5
can be contaminated by bright Ic SNe like 1992ar.

With current detectors and telescopes, which 
make detection of the SiII \l6355 line difficult for SNe with z$\gapprox$0.5,
the most reliable means of eliminating
bright Type~Ic SNe from the high~z SNe samples
SNe will be:
(1) To obtain high S/N ratio spectra in the visual and red regions in order
to detect the typically much weaker SiII lines at 4130 and 5051 \AA\  (note
that the line SiII \l5972 could appear merged with NaI~D, so its diagnostic
power is limited), and the SII lines at 5468, 5612 and 5654 \AA,
which have never been claimed to be
present in the spectrum of Type~Ic SNe, 
{\em and},
(2) to obtain well time sampled two color light curves with good signal-to-noise ratio
including at least a pre--maximum point and, especially, a couple of good
points, with photometric uncertainty smaller than $\sim$0.05 mag
on the early exponential tail, between 40 and 60 days after maximum light.

A good S/N ratio data
set for a high~z SN would allow to apply the following criteria.
If a SN has a red color, does not show evidence for
SiII \l4130, SiII \l5051, SII \l5468, SII \l5612, and SII~\l5654 lines,
displays a maximum noticeably narrower than that of a Type~Ia SN, {\em or}
a light curve which remains brighter than the Type~Ia SNe light curve
templates on the early exponential tail,
it should be suspect of being a Type Ic SN and excluded from the samples.

How could these luminous Type Ic SNe affect the high z Hubble diagrams if not
removed from the samples? 
According to the current, scarce, database, and
opposite to what Type~Ia SNe display, the shapes of
light curves of Transition Type SNe do not seem to correlate with their
intrinsic brightness.
As we have shown,
the same type of light curves fit SNe with very different intrinsic brightness.
In the cases of a slow Type~Ic SN fit for SN~1992ar this means
M$_{\rm V}= -17.4$ (SN 1983N), and M$_{\rm V}= -19.2$ (SN 1992ar),
while in the case of a fast Type~Ic SN fit for SN~1992ar it means
M$_{\rm V}= -18.1$ (SN 1994I), and M$_{\rm V}= -20.2$ (SN 1992ar).
It could well be that the process which determines the amount of \Ni\  ejected
in core--collapse explosions is intrinsically
chaotic and bears no relation to the mass to energy ratio of the explosions,
responsible for the light curve shapes.
It would appear, in principle,
that the main effect of not removing objects like
SN~1992ar from the high~z Type Ia SNe samples will be
an increase in the scatter of the final Hubble Diagrams.

On the other hand, systematic shifts may enter through
Malmquist bias.
If most of the Type~Ic which go unchecked and are included
in the samples are as bright as SN 1992ar (i.e.
slightly brighter than a typical Type Ia SN), then
the sample of distant SNe will be biased in the
sense of appearing brighter than they should.
The result of this will be that the distances estimated
will be shorter than they are, and
consistent with a positive deceleration parameter q$_0$, or
alternatively, with higher values of $\Omega_{\rm M}$ and lower values
of $\Omega_\Lambda$.

\acknowledgements 
Support for AC, at different stages of this project,
was provided by Fundaci\'on Antorchas Argentina under project
A-13313, by P. Universidad Cat\'olica de Chile under DIPUC Project 97/12E,
and by FONDECYT, Chile, under Project 1980803.
NBS and MMP acknowledge the support from NASA through grants GO-2563.01-87A
and GO-6020 from STScI.

\appendix
\section{A note on SN~1998bw in ESO 184-G82}

While this paper was in the final stages of writing and submission the first
results on SN~1998bw became available (Saddler et al. 1998,
Kulkarni et al., 1998, Galama et al. 1998).
Although a detailed comparison of SNe 1998bw and 1992ar is beyond the scope of
the present paper, a note cross linking them is necessary.
SN~1998bw was discovered well within the error box of GRB 980425 and there is
compelling evidence that it was associated with the burst.
SN~1998bw displayed peculiarities in its light curve and spectra, with a very
fast rise to maximum (Woosley, Eastman \& Schmidt 1999), and photospheric
expansion velocities of $\sim$28,000 km s$^{-1}$ ten days after explosion
(Iwamoto et al. 1998).
SN~1998bw was also recognized as a very bright event.
At an absolute magnitude of M$_{\rm V}= -19.35\pm0.05$ (Galama et al. 1998),
however, it could have been dimmer than SN~1992ar.
If the two events correspond to core--collapse explosions, they show
very clearly that this mechanism does not result in a correlation between
\Ni\  mass ejection and mass to energy ratio in the ejecta.

\newpage
\figcaption{\label{fi:chart}
Chart of the region of SN 1992ar.
The sequence of local standard stars used to calibrate the brightness of the
SN is labeled by numbers positioned towards the NE of the referenced stars.
Note that objects number 5 and 9 are not stars but galaxies.
The three galaxies nearby the SN are labeled by letters.
The position of the SN is given by a white square dot towards the SE of these
group of galaxies.
}

\figcaption{\label{fi:spec1}
Comparison of the spectrum of SN~1992ar with those of the Type~Ia SN~1989B
(Wells et al. 1994) and 1991T (Phillips et al. 1992).
The wavelength scales of the spectra have been corrected for the redshift of
the parent galaxies.
The phase of the spectrum of SN~1992ar corresponds to the 
slow Type~Ic light curve fit discussed in \S~6.2.
Identification of the main features, labeled by lower case
letters above the spectrum of SN~1991T are:
(a) CaII H \& K,
(b) FeII ``\l4274'',
(c) FeII ``\l4555'',
(d) FeII \l5018,
(e) FeII \l5169,
(f) NaI~D / SiII ``\l5972,''
(g) SiII ``\l6355.''
Secondary weaker features which could be used to distinguish Type~Ia from
Type~Ic SNe are labeled with capital letters, as follows:
(A) SiII \l4130,
(B) SiII \l 5051,
(C) SII ``\l5468,''
and,
(D) SII ``\l\l5612,'' ``5654.''
}

\figcaption{\label{fi:spec2}
Comparison of the spectrum of SN~1992ar with those of the Type~Ic SN~1983V
(Clocchiatti et al. 1997) and 1987M (Filippenko, Porter, \& Sargent 1990).
The wavelength scales of the spectra have been corrected for the redshift of
the parent galaxies.
The phase of the spectrum of SN~1992ar corresponds to the
slow Type~Ic light curve fit discussed in \S~6.2.
Identification of the main features, labeled by letters above the spectrum of
SN~1983V are:
(a) CaII H \& K,
(b) FeII ``\l4274'',
(c) FeII ``\l4555'',
(d) FeII \l5018,
(e) FeII \l5169,
(f) NaI~D,
(g) SiII ``\l6355.''
}

\figcaption{\label{fi:specg}
Spectra of the three galaxies nearby SN~1992ar.
The wavelength scale of the spectra are given in the observer's frame
(i.e. not corrected by redshift of the parent galaxies).
The forbidden emission lines are marked by vertical lines and labeled on the
top spectrum (Galaxy A).
They are, in increasing order of wavelength,
[O II] \l\l 3726,3729;
[O III] \l\l 4959,5007;
and
[NI] \l 5198.
The hydrogen recombination lines \Hd, \Hg, and \Hb, with the latter seen in
emission, have been also marked on this top spectrum.
For clarity of the plot, the hydrogen recombination lines \He, \Hx, and the
CaII H\&K lines, all seen in absorption, have been marked on the bottom
spectrum (Galaxy C).
}

\figcaption{\label{fi:Kcorr}
The K--corrections in V for Type~Ic slow, Type~Ic fast, and Ia SNe at z$=$ 0.145.
The vertical scale of both panels is the same to facilitate comparisons.
}

\figcaption{\label{fi:lcexample}
Observed raw photometry of SN~1992ar (with no K corrections applied),
compared with the
different final light curves that result when different K--correction
functions are assumed.
Solid circles, joined by solid line, show the raw photometry (V$_{0.145}$).
Open circles joined by a dotted line, correspond to V$_{0.145}$ minus the
K--correction of a slow Type~Ic SN.
Open squares joined by a short dashed line correspond to V$_{0.145}$ minus the
K--correction of a fast Type~Ic SN.
Open triangles joined by a long dashed line correspond to V$_{0.145}$ minus the
K--correction of a Type~Ia SN.
A phase of 3 days after V maximum was assumed for all light curves.
The arrow marks the time position of the spectrum.
}

\figcaption{\label{fi:LCIa1}
Type~Ia SNe light curves compared with the CCD photometry of SN~1992ar.
The top panel shows the fitting of a slow decaying luminous Type~Ia SN,
illustrated in this case by SN~1992bc (Hamuy et al. 1996).
The RMS of the fit is minimized when a phase of 3 days after V maximum is
assumed for the first CCD observation.
The bottom panel shows the fitting of a more typical Type~Ia SN, exemplified
by SN~1992A.
The RMS of the fit is minimized when a phase of 6 days after V maximum is
assumed for the first CCD point.
}

\figcaption{\label{fi:LCIa2}
Fitting of the light curve of SN~1992A to the observed photometry of SN~1992ar
assuming that the phase of the first CCD observation is $\sim$17 days after maximum
light in V.
The arrow marks the time position of the spectrum.
The open squares with large error bars are the rest frame V magnitudes estimated
from the plate, using two different estimates of the color of the SN
(see Table~\protect{\ref{ta:mphot}}).
}

\figcaption{\label{fi:LCIc}
Fitting of Type~Ic SNe light curves to the observed photometry of SN~1992ar.
The top panel shows the fit of a fast Type~Ic SN, exemplified by SN~1994I
(Richmond et al. 1996).
The arrow marks the time position of the spectrum.
The bottom panel shows the fitting of SN~1993J, whose light curve is
a good match to those of slow Type~Ic SN.
The arrow marks the time position of the spectrum.
In both panels, the open square with a large error bar shows the rest frame
V magnitude estimated from the discovery plate
(see Table~\protect{\ref{ta:mphot}}).
}

\newpage

\begin{table}

\caption{ Photometric Observations of SN~1992ar \label{ta:CCDobs}}

\begin{tabular}{rcccl}

\tableline
\tableline

  Date\tablenotemark{a} &   V            &   B                  &  R        & Source \\

\tableline

 35.93 & 20.11 $\pm$ 0.04 & ---              & ---       & CTIO 0.91m + CCD --- R. Avil\'es/C. Smith \\
 36.78 & ---                & 21.05 $\pm$ 0.20 & ---       &  B$-$V=0.85 (spectrum) and V=20.20.\\
 40.80 & ---                & 21.47 $\pm$ 0.19 & ---       &  CTIO 0.91m + CCD --- E.P. Rubenstein \\
 40.81 & 20.54 $\pm$ 0.06 & ---              & ---       & CTIO 0.91m + CCD --- E.P.  Rubenstein \\
 47.71 & 21.22 $\pm$ 0.12 & ---              & ---       & CTIO 0.91m + CCD --- R. Avil\'es \\
 67.55 & 21.91 $\pm$ 0.05 & ---              & 21.14 $\pm$ 0.07 & ESO 3.6m + EFOSC1 --- E. Cappellaro \\
 98.78 &  ---             & ---              & 22.13 $\pm$ 0.07 & ESO/MPI 2.2m + EFOSC2 --- M. Della Valle\\
 98.79 & 22.31 $\pm$ 0.07 & ---              & ---       & ESO/MPI 2.2m + EFOSC2 --- M. Della Valle\\
 99.65 & 22.30 $\pm$ 0.08 & ---              & ---       & ESO/NTT + EMMI --- M. Della Valle \\
156.54 & 23.22 $\pm$ 0.15 & ---              & ---       & ESO 3.6m + EFOSC1 --- M. Della Valle \\

\end{tabular}
\tablenotetext{a}{JD$-$2448800}

\end{table}

\begin{table} [f]

\caption{Final Epoch Images of SN~1992ar Region \label{ta:FEImages}}

\begin{tabular}{lrrll}

\tableline
\tableline

  Date &   Phase  &  Filter  & Telescope and Instrument & Observer \\

\tableline

  1993, Sep 9 - 6:35  &   409    &    B     & CTIO 4.0m -- Cass CCD Camera & M. Navarrete \\ 
  1993, Nov 11 - 3:53  &   472    &    V     & CTIO 4.0m -- Cass CCD Camera & R. Avil\'es \\
  1996, Dec 15 - 2:21  &   1602    &    R     & ESO 2.2m  -- EFOSC 2         & C. Zanin 
\end{tabular}

\end{table}

\begin{table}

\caption{ Local Standard Sequence \label{ta:STD}}

\begin{tabular}{lcccc}

\tableline
\tableline

   ID    & V                &  B           & B$-$V          &  R          \\
\tableline
 92ar-1  & 16.15 $\pm$ 0.02 & 16.79 $\pm$ 0.02 & 0.65 $\pm$ 0.03 & 15.67 $\pm$ 0.04 \\ 
 92ar-2  & 18.48 $\pm$ 0.02 & 19.37 $\pm$ 0.03 & 0.89 $\pm$ 0.03 & 17.83 $\pm$ 0.05 \\ 
 92ar-3  & 16.72 $\pm$ 0.02 & 17.85 $\pm$ 0.03 & 1.14 $\pm$ 0.03 & 15.87 $\pm$ 0.07 \\ 
 92ar-4  & 17.61 $\pm$ 0.02 & 19.05 $\pm$ 0.03 & 1.44 $\pm$ 0.03 & 16.42 $\pm$ 0.09 \\ 
 92ar-5  & 17.53 $\pm$ 0.02 & 18.81 $\pm$ 0.03 & 1.28 $\pm$ 0.03 &  --- \\ 
 92ar-6  & 17.05 $\pm$ 0.02 & 17.71 $\pm$ 0.03 & 0.67 $\pm$ 0.03 & 16.57 $\pm$ 0.04 \\ 
 92ar-7  & 17.99 $\pm$ 0.02 & 18.61 $\pm$ 0.03 & 0.62 $\pm$ 0.03 &  --- \\ 
 92ar-8  & 20.73 $\pm$ 0.05 & 21.60 $\pm$ 0.12 & 0.87 $\pm$ 0.13 & 19.95 $\pm$ 0.06 \\ 
 92ar-9  & 18.31 $\pm$ 0.02 & 19.70 $\pm$ 0.03 & 1.39 $\pm$ 0.03 &  --- \\ 
 92ar-11 & 20.04 $\pm$ 0.03 & 21.41 $\pm$ 0.09 & 1.38 $\pm$ 0.10 & 18.70 $\pm$ 0.08 \\ 
 92ar-12 & 19.97 $\pm$ 0.03 & 20.23 $\pm$ 0.05 & 0.26 $\pm$ 0.06 & 20.03 $\pm$ 0.03 \\ 
 92ar-13 & 20.22 $\pm$ 0.03 & 21.66 $\pm$ 0.11 & 1.44 $\pm$ 0.11 & 18.97 $\pm$ 0.09  

\end{tabular}

\end{table}

\begin{table}

\caption{Rest Frame Magnitudes at Discovery \label{ta:mphot}}

\begin{tabular}{lrrrrrrrrr}
\tableline
\tableline
Type & Ph.\tablenotemark{a} & K$_{\rm IIaO}$ & \multicolumn{1}{c}{K$_{\rm B}$} & \multicolumn{1}{c}{K$_{\rm V}$} & \multicolumn{1}{c}{B-V$_{0.145}$} & \multicolumn{1}{c}{B-V$_0$} & \multicolumn{1}{c}{B$_0$} & \multicolumn{1}{c}{V$_0$}  \\
\tableline
Ia\tablenotemark{b}      &   12 & 0.55 & 0.38$\pm 0.1$ &  0.03$\pm 0.05$ & 0.80$\pm 0.22$ & 0.45$\pm 0.20$ & 19.83$\pm 0.26$ & 19.38$\pm 0.34$ \\
Ia\tablenotemark{c}      &   12 & 0.55 & 0.38$\pm 0.1$ &  0.03$\pm 0.05$ & 0.40$\pm 0.20$ & 0.05$\pm 0.22$ & 19.91$\pm 0.26$ & 19.86$\pm 0.34$ \\
Ic Slow\tablenotemark{b} & -1.5 & 0.43 & 0.16$\pm 0.1$ & -0.05$\pm 0.05$ & 0.38$\pm 0.22$ & 0.20$\pm 0.20$ & 20.07$\pm 0.26$ & 19.90$\pm 0.34$ \\
Ic Slow\tablenotemark{c} & -1.5 & 0.43 & 0.16$\pm 0.1$ & -0.05$\pm 0.05$ & 0.56$\pm 0.20$ & 0.35$\pm 0.22$ & 20.04$\pm 0.26$ & 19.69$\pm 0.34$ \\
Ic Fast\tablenotemark{b} &  2.5 & 0.59 & 0.25$\pm 0.1$ &  0.03$\pm 0.05$ & 0.77$\pm 0.22$ & 0.55$\pm 0.20$ & 19.79$\pm 0.26$ & 19.24$\pm 0.34$ \\
Ic Fast\tablenotemark{c} &  2.5 & 0.59 & 0.25$\pm 0.1$ &  0.03$\pm 0.05$ & 0.64$\pm 0.20$ & 0.42$\pm 0.22$ & 19.81$\pm 0.26$ & 19.39$\pm 0.34$
\end{tabular}
\tablenotetext{a}{Phase in rest frame days after V maximum.}
\tablenotetext{b}{Rest frame color at discovery assumed.}
\tablenotetext{c}{Rest frame color at discovery extrapolated from the spectrum.}

\end{table}

\begin{table}

\caption{Rest Frame Light Curve of SN~1992ar (Ic fits)\label{ta:Icfits}}

\begin{tabular}{rrccrcc}
\tableline
\tableline
       & \multicolumn{3}{c}{Slow Type Ic SN Fit} & \multicolumn{3}{c}{Fast Type Ic SN Fit} \\
\tableline
Date\tablenotemark{a}   &  Phase & K$_{\rm V, Slow}$ & V$_{\rm Slow}$ & Phase & K$_{\rm V, Fast}$ & V$_{\rm Fast}$ \\
\tableline
35.93  &   3.00 &  0.04   & 20.07 $\pm$ 0.09 & 7.000 &  0.11 & 20.00 $\pm$ 0.08 \\
40.81  &   7.26 &  0.12   & 20.43 $\pm$ 0.11 & 11.26 &  0.17 & 20.37 $\pm$ 0.09 \\
47.71  &  13.29 &  0.20   & 21.02 $\pm$ 0.14 & 17.29 &  0.13 & 21.09 $\pm$ 0.14 \\
67.55  &  30.61 &  0.28   & 21.63 $\pm$ 0.10 & 34.61 & -0.21 & 22.12 $\pm$ 0.09 \\
98.79  &  57.90 &  0.10   & 22.22 $\pm$ 0.11  & 61.90 & -0.29 & 22.60 $\pm$ 0.10 \\
99.65  &  58.65 &  0.10   & 22.20 $\pm$ 0.12 & 62.65 & -0.29 & 22.59 $\pm$ 0.11 \\
156.54 & 108.33 & -0.05   & 23.27 $\pm$ 0.17 & 112.33 & -0.43 & 23.65 $\pm$ 0.17
\end{tabular}
\tablenotetext{a}{JD$-$2448800}

\end{table}

\begin{table}[f] \label{ta:absmag}
\caption{M$_{\rm V}$ of Transition Type SNe}

\begin{tabular}{lcl}
\tableline
\tableline
SN     &    M$_{\rm V}$       & Source  \\
\tableline
1983N  &    -17.4  $\pm$ 0.2  & Clocchiatti et al. (1996) \\
1983V  &    -18.1  $\pm$ 0.2  & Clocchiatti et al. (1997) \\
{\bf 1992ar} &  -19.2 $\pm$ 0.2 & This paper\tablenotemark{a} \\
{\bf 1992ar} &  -20.2 $\pm$ 0.2 & This paper\tablenotemark{b} \\
1993J  &    -17.7 $\pm$ 0.1 & Table~5 in Clocchiatti et al. (1997) \\
1994I  &    -18.1  $\pm$ 0.6  & Richmond et al. (1996) \\
\end{tabular}
\tablenotetext{a}{If it was a Slow Type Ic SN. Uncertainty estimated from the template fitting.}
\tablenotetext{b}{If it was a Fast Type Ic SN. Uncertainty estimated from the template fitting.}

\end{table}

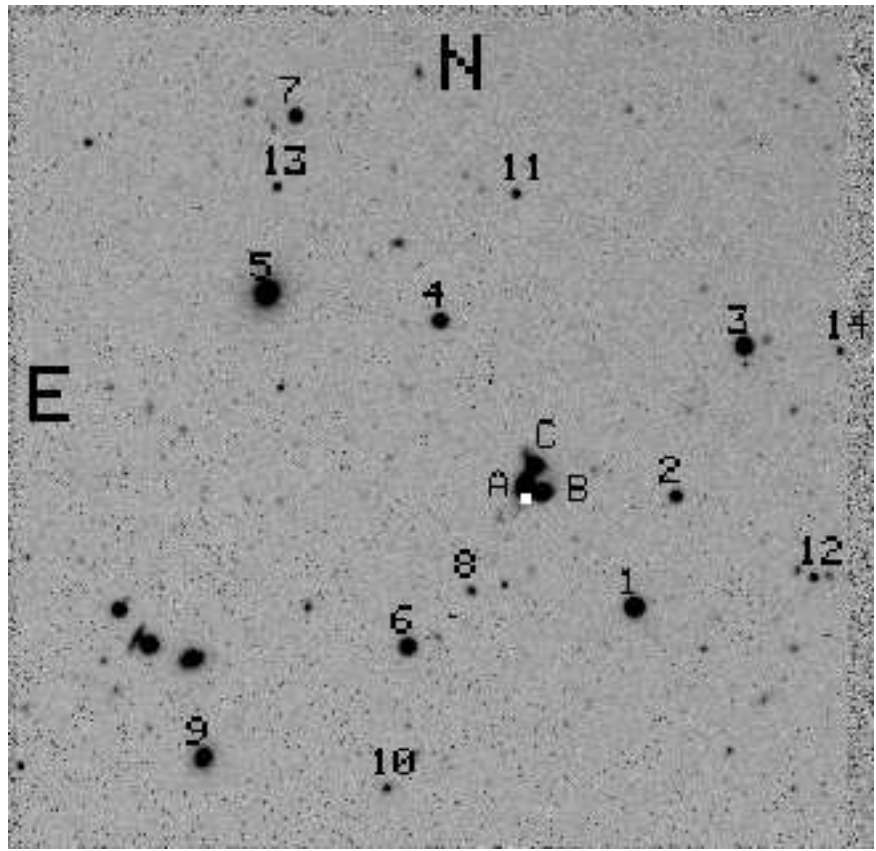
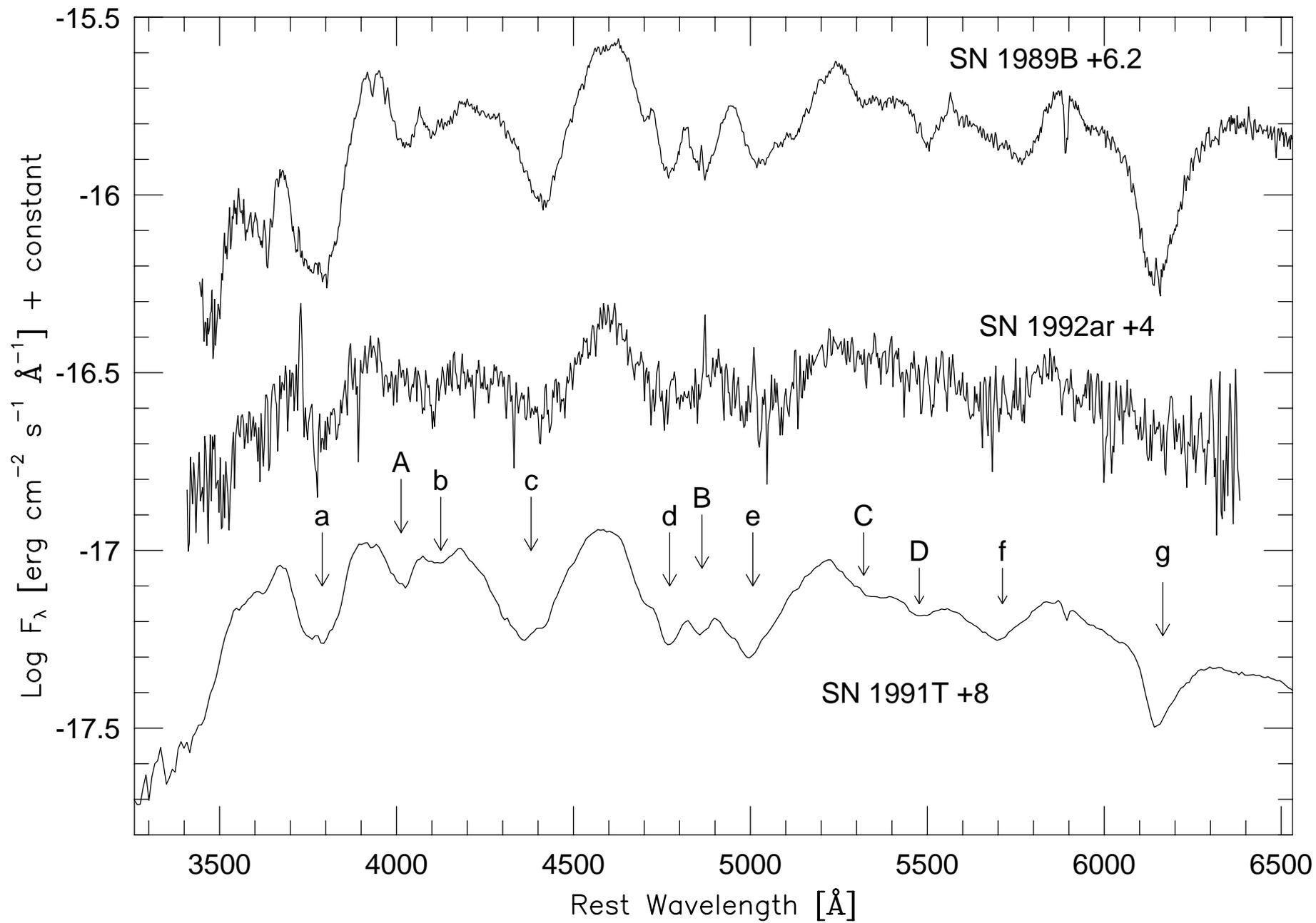
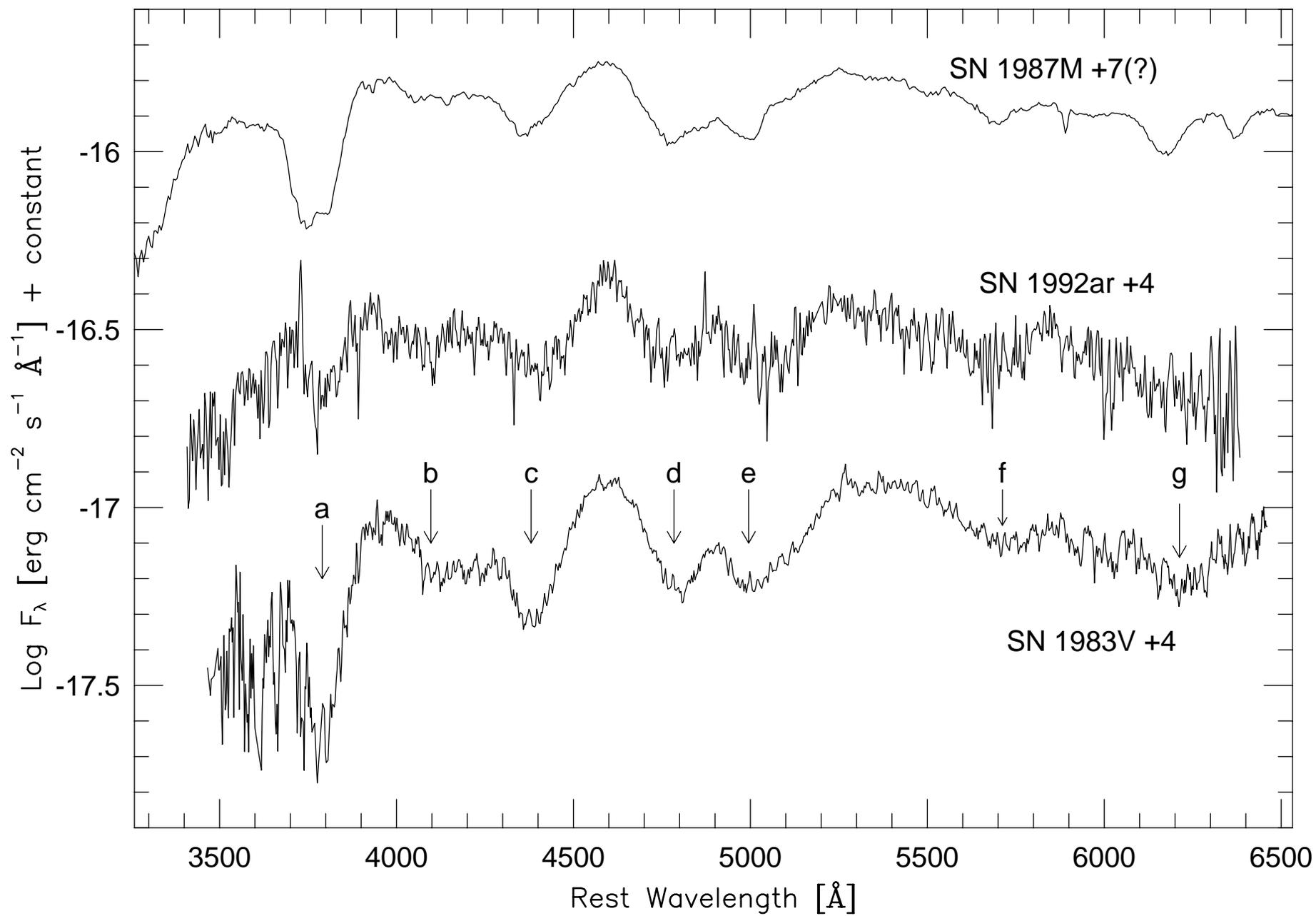
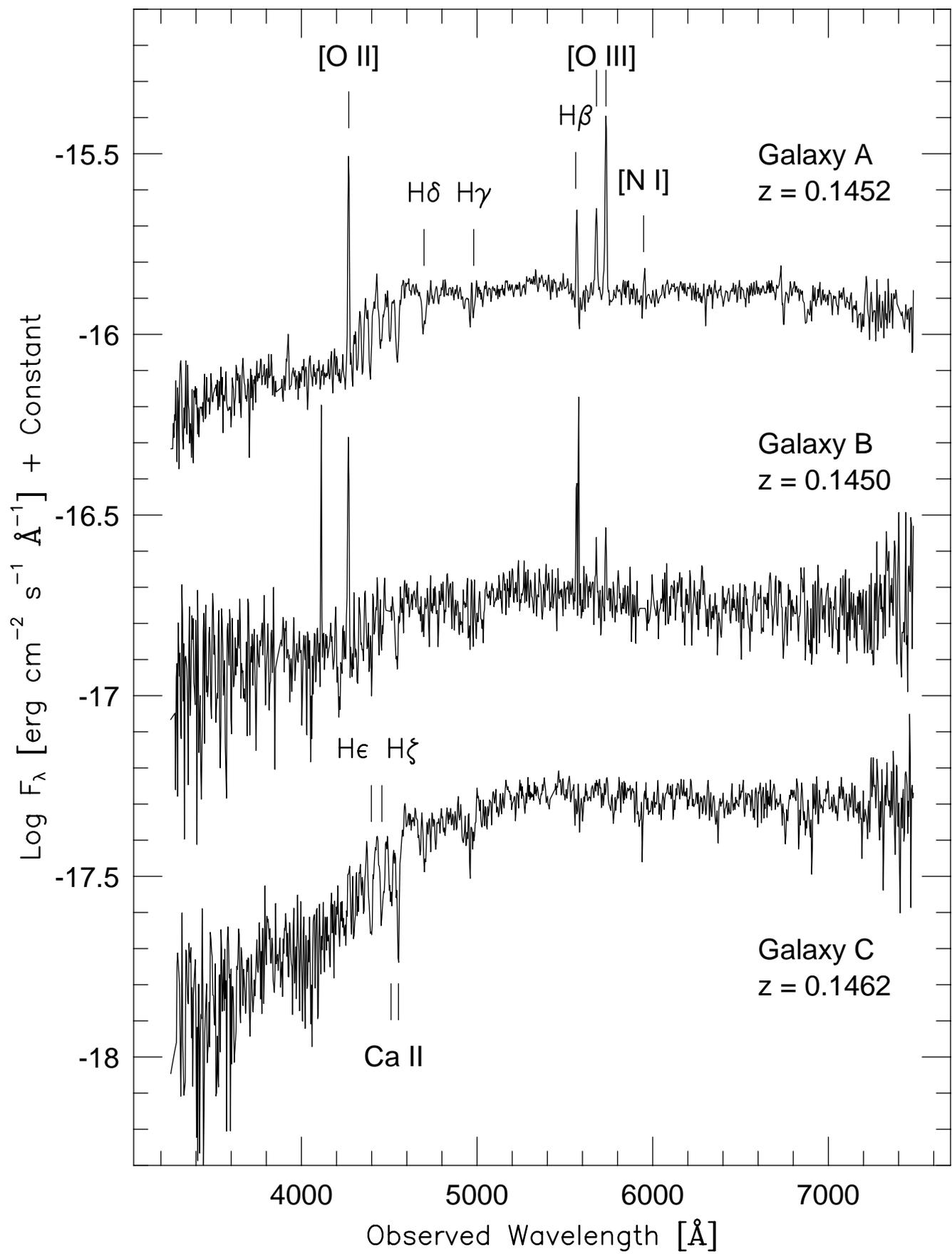
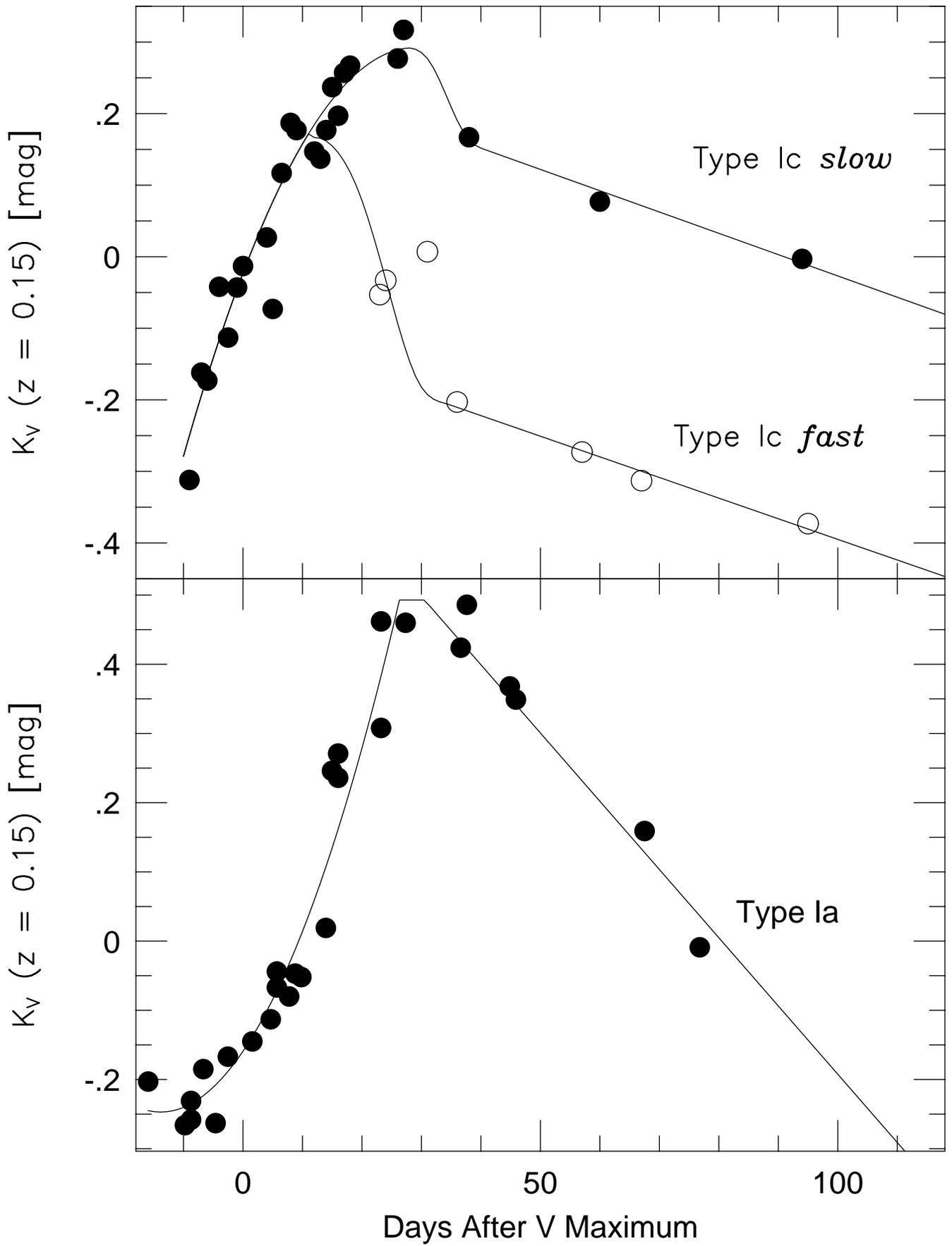
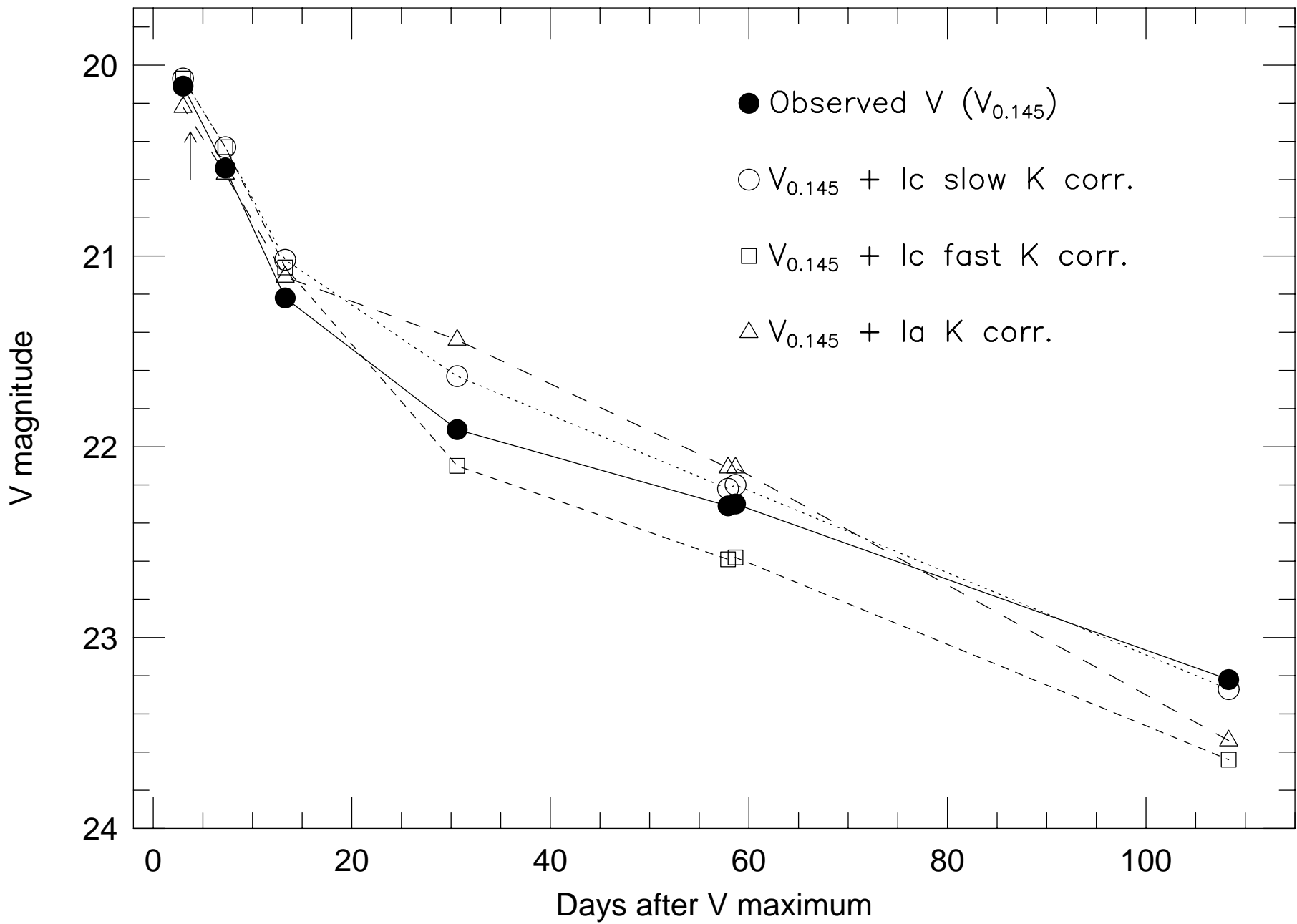
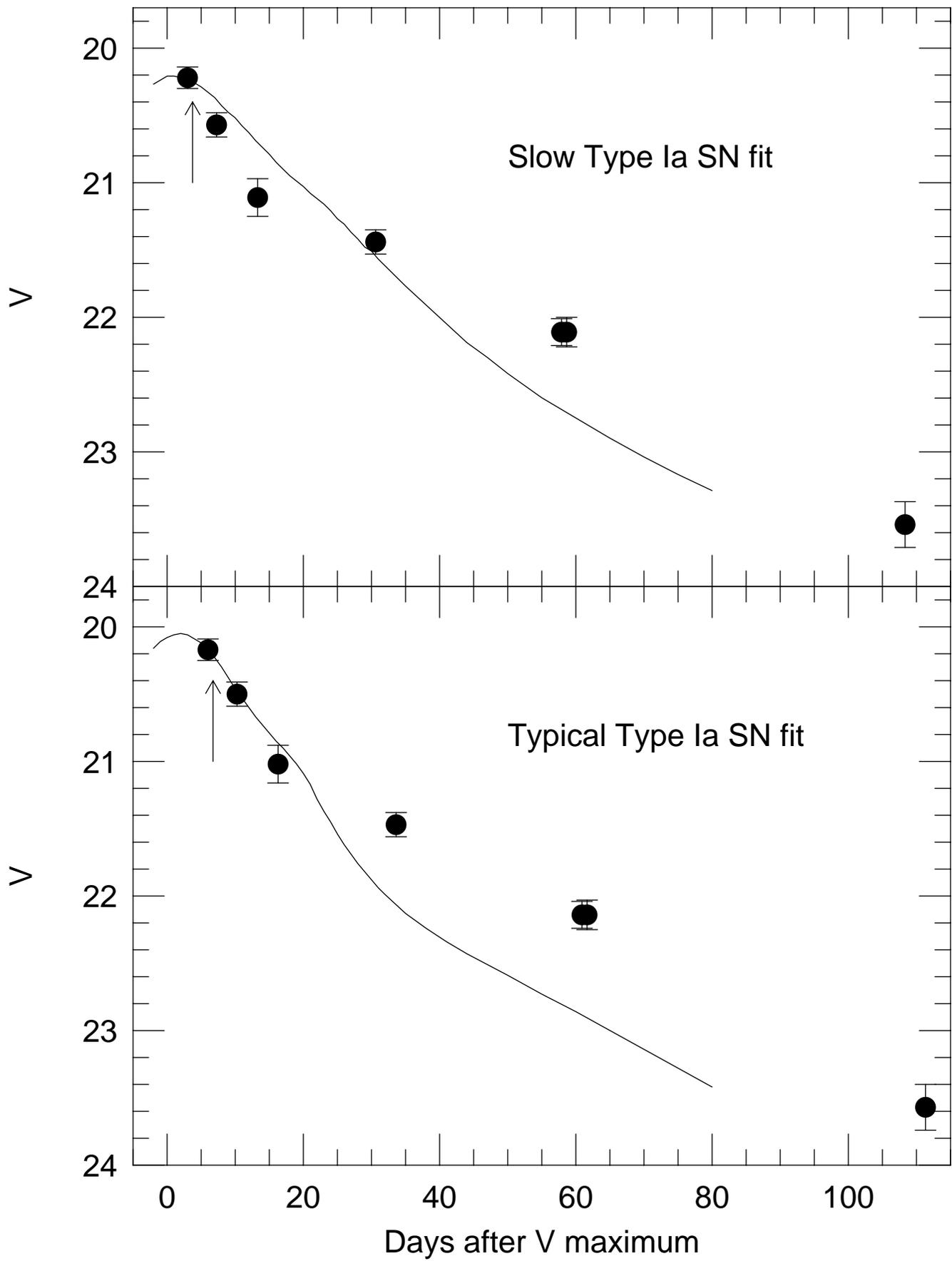
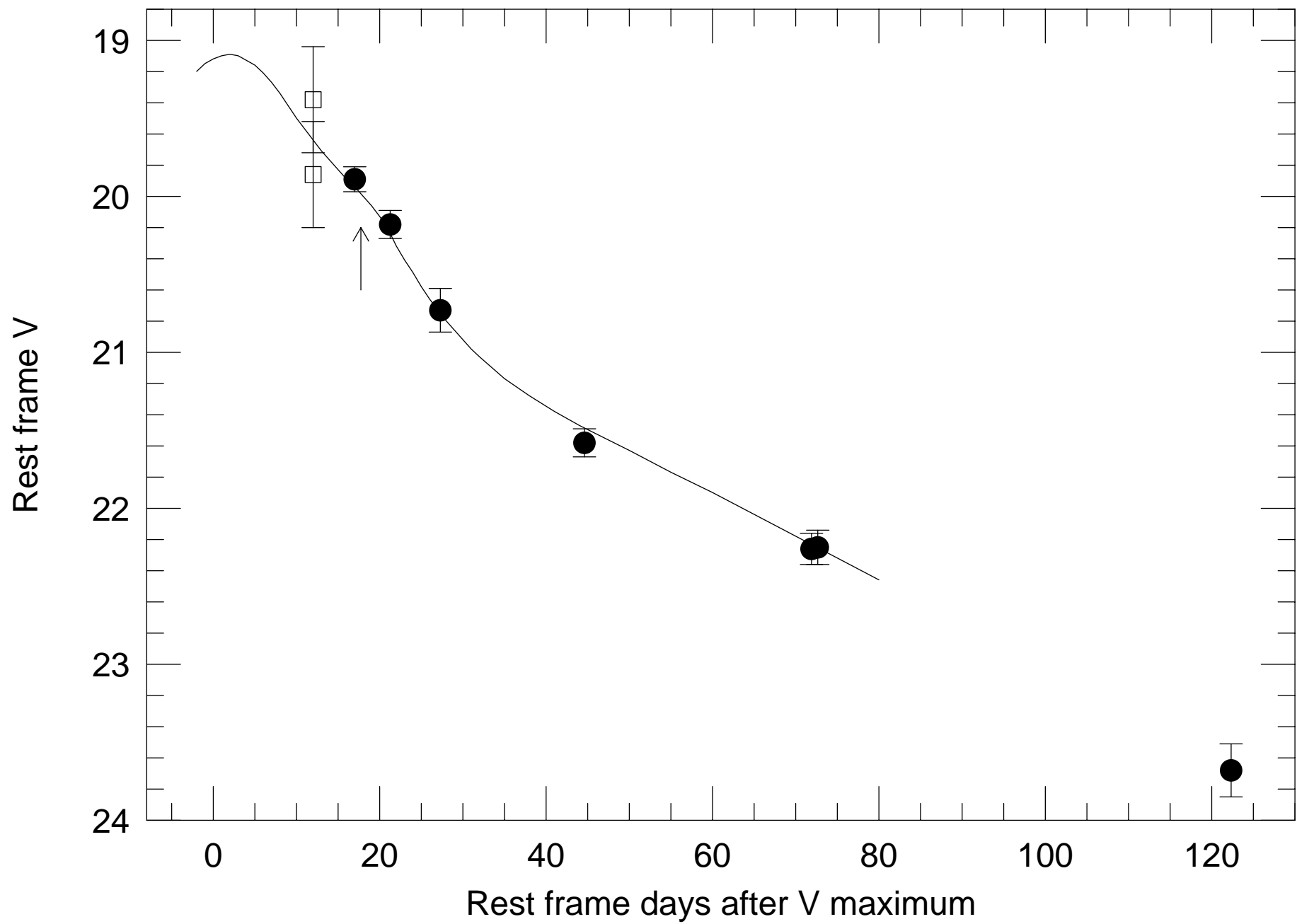
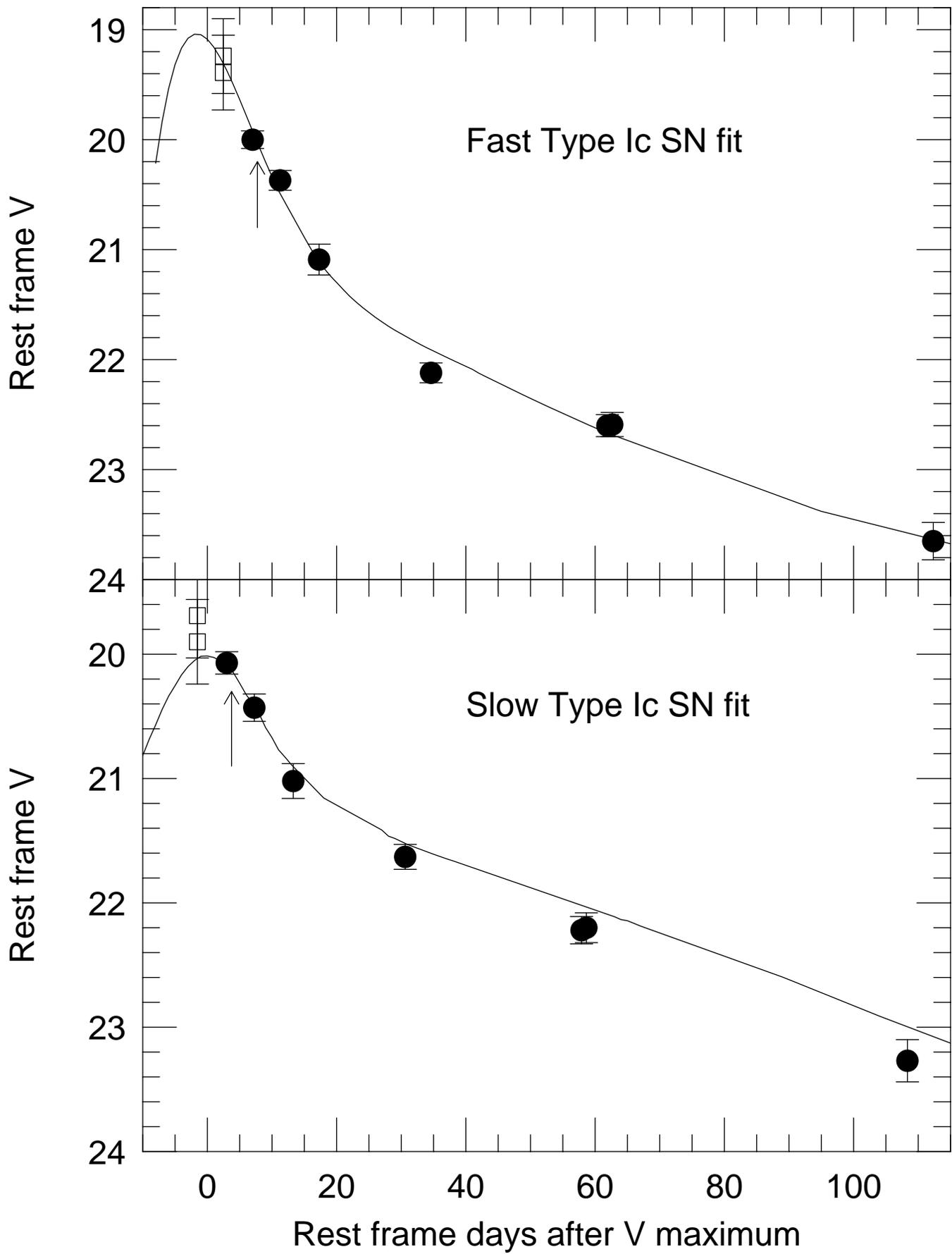
\end{document}